\documentclass[aps,pra,reprint,groupedaddress]{revtex4-1}
\usepackage{graphicx}
\usepackage{color}
\usepackage{transparent}
\graphicspath{{figure/}}
\begin{document}

%\preprint{}

\title{Microwave-stimulated Raman adiabatic passage in a Bose-Einstein condensate on an atom chip}


\author{M. Dupont-Nivet$^{1,2}$, M. Casiulis$^{1}$, T. Laudat$^{1}$, C. I. Westbrook$^{2}$ and S. Schwartz$^{1}$}
%\thanks{}
\affiliation{${}^{1}$Thales Research and Technology France, Campus Polytechnique, 1 av. Fresnel, 91767 Palaiseau, France \\
${}^{2}$Laboratoire Charles Fabry de l'Institut d'Optique, Campus Polytechnique, 2 av. Fresnel, 91127 Palaiseau, France}

\date{\today}

\begin{abstract}
We report the achievement of stimulated Raman adiabatic passage (STIRAP) in the microwave frequency range between internal states of a Bose-Einstein condensate (BEC) magnetically trapped in the vicinity of an atom chip. The STIRAP protocol used in this experiment is robust to external perturbations as it is an adiabatic transfer, and power-efficient as it involves only resonant (or quasi-resonant) processes. Taking into account the effect of losses and collisions in a non-linear Bloch equations model, we show that the maximum transfer efficiency is obtained for non-zero values of the one- and two-photon detunings, which is confirmed quantitatively by our experimental measurements.
\end{abstract}


\maketitle

%%%%%%%%%%%%%%%%%%%%%%%%%%%%%%%%%%%%%%%%%%%%%%%%%%%%%%%%%%%%%
%                                                           %
%%%%%%%%%%%%%%%%%%%%%%%%%%%%%%%%%%%%%%%%%%%%%%%%%%%%%%%%%%%%%
\section{Introduction}

Since its first demonstration as an efficient tool for exciting vibrational states in molecular beams \cite{Gaubatz1988,Gaubatz1990}, stimulated Raman adiabatic passage (STIRAP) \cite{Shore1991,Bergmann1998,Vitanov2001} has been applied to various situations involving multi-level systems \cite{Goldner1994}, such as quantum information processing \cite{Kis2002,Timoney2011,Webster2013}, atomic physics \cite{Pillet1993,Weitz1994b,Nandi2004,Graefe2006,Rab2008,Snigirev2012}, cold molecules \cite{Drummond2002,Winkler2005,Itin2007,Kuznetsova2009} and deterministic single-photon source \cite{Kuhn2002}. As regards theoretical studies, an extensive literature exists as well  \cite{Kuklinski1989,Carroll1990,Fewell1997,Vitanov1997b,Vitanov1997,Romanenko1997}. An important advantage of this technique is that it permits quasi-resonant multi-photon transitions without populating unstable intermediate states. Moreover, as an adiabatic transfer method, it is relatively insensitive to fluctuations in the experimental parameters \cite{Bergmann1998}.

In this paper, we demonstrate the use of STIRAP with two microwave frequencies to perform population transfer between hyperfine ground states of a ${}^{87}$Rb Bose-Einstein condensate (BEC) trapped in the vicinity of an atom chip. Our STIRAP starts from the highest energy level in an upside-down $\Lambda$-system \cite{Shore2013}. Because of collision losses within the BEC in the target state, the maximum transfer efficiency is obtained for non-resonant driving fields. This in quantitatively confirmed by a theoretical model based on non-linear Bloch equation \cite{Shore1990}. 

More precisely, we prepare the BEC in the $\left|F=2,m_F=2\right>$ hyperfine level of the $5^2S_{1/2}$ ground state and transfer it coherently into $\left|F=2,m_F=1\right>$, from which the two-photon transition $\left|F=2,m_F=1\right> \longleftrightarrow \left|F=1,m_F=-1\right>$, which has very good coherence properties \cite{Harber2002,Treutlein2004,Bohi2009,Deutsch2010}, can be addressed. Preparing the BEC in $\left|F=2,m_F=2\right>$, whose magnetic moment is twice as big as other trappable hyperfine levels of $5^2S_{1/2}$, has the advantage of bigger magnetic forces for equivalent electrical power dissipation, allowing for example to capture atoms from a magneto-optical trap farther away from the current-carrying wires \cite{Farkas2010,Huet2012}, or to increase the trap confinement during evaporative cooling \cite{Ketterle1996b}.

\begin{figure}
\centering \def\svgwidth{0.35\textwidth}
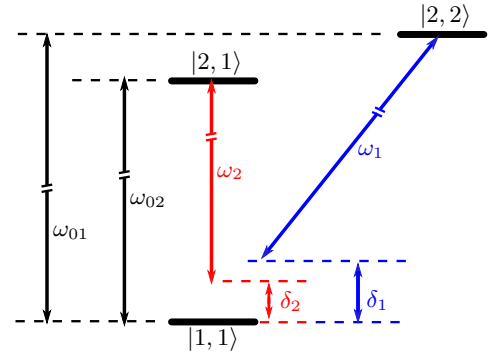
\caption{\label{fig_STIRAP_levels} Hyperfine ground states of $^{87}$Rb involved in the STIRAP process, forming a upside-down $\Lambda$-system. We transfer population from state $\left|F=2,m_F=2\right>$ to state $\left|F=2,m_F=1\right>$ via the unpopulated intermediate level $\left|F=1,m_F=1\right>$. We write~: $\left|F=2,m_F=2\right> = \left|e_1\right>$, $\left|F=2,m_F=1\right> = \left|e_2\right>$ and $\left|F=1,m_F=1\right> = \left|g\right>$. The one- and two-photon detunings are respectively defined as $\Delta\equiv(\delta_1+\delta_2)/2$ and $\delta\equiv\delta_1-\delta_2$.}
\end{figure}

In order to benefit from these advantages, a reliable protocol for coherent population transfer between $\left|F=2,m_F=2\right>$ and $\left|F=2,m_F=1\right>$ is required. The most straightforward transfer method would be to use a single-photon transition in the radio-frequency range. However, with the typical magnetic field values used in our experiment (less than a few tens of Gauss), the Zeeman effect is nearly linear \cite{Steck2003}, and single photon transitions would spread the atoms in all (equally spaced) Zeeman sub-levels of the $F=2$ manifold. A second possibility is to use a two-photon transition in the microwave range, with $\left|F=1,m_F=1\right>$ as an intermediate level. For such a transition, the two-photon detuning ($\delta\equiv\delta_1-\delta_2$ in figure \ref{fig_STIRAP_levels}) could be arbitrarily low to maximize the transfer efficiency, but the one-photon detuning $\Delta\equiv(\delta_1+\delta_2)/2$ has to remain much bigger than $\Omega_1$ and $\Omega_2$ (which are the maximum available Rabi frequencies associated with the two microwave fields drawn in figure \ref{fig_STIRAP_levels}) in order to keep the population of the anti-trapping state $\left|F=1,m_F=1\right>$ at a reasonably low level \cite{Gentile1989}. When the latter condition is fulfilled, the two-photon transition is equivalent to a single-photon transition with the effective Rabi frequency $\Omega_\textrm{\scriptsize eff}=\Omega_{1}\Omega_{2}/(2|\Delta|)$ \cite{Gentile1989}. It can thus be used to implement a population transfer either with a $\pi$ pulse (though it would imply more stringent constraints on the stability of experimental parameters) \cite{Shore1992} or with an adiabatic passage across resonance (which then has to be much slower than the time scale $1/\Omega_\textrm{\scriptsize eff}$). In comparison, the STIRAP protocol which we consider in this paper allows us to perform the same adiabatic population transfer on a shorter time scale, on the order $1/\sqrt{\Omega_{1}^2+\Omega_{2}^2}$.

This article is organized as follows. In section \ref{sec_theoritical_framework}, we briefly summarise the basic principles of STIRAP, and describe how the usual theoretical model is adapted to account for losses and collisions which play a major role in our experiment. Section \ref{sec_experiment} is devoted to the description of our experimental protocol and results. In section \ref{sec_Simulation}, we report numerical simulations of the STIRAP process, in good quantitative agreement with our measurements. Finally, we physically discuss in section \ref{sec_discution} the need for non-zero one- and two-photon detunings for maximum transfer efficiency.

%%%%%%%%%%%%%%%%%%%%%%%%%%%%%%%%%%%%%%%%%%%%%%%%%%%%%%%%%%%%%
%                                                           %
%%%%%%%%%%%%%%%%%%%%%%%%%%%%%%%%%%%%%%%%%%%%%%%%%%%%%%%%%%%%%
\section{Theoretical framework}
\label{sec_theoritical_framework}

\subsection{Basics of the STIRAP protocol}
\label{sec_basics}

Let us consider the three-level upside-down $\Lambda$ system of figure \ref{fig_STIRAP_levels}, with all the atoms initially in the highest energy level $\left|e_{1}\right>$. The two levels $\left|e_{1,2}\right>$ are coupled to $\left|g\right>$ by two microwave fields with time-dependent Rabi frequency $\Omega_{1,2}(t)$ and  constant frequency $\omega_{1,2}$ detuned by $\delta_{1,2} = \omega_{1,2} - \omega_{0\,1,2}$, where $\hbar \omega_{0\,1,2}$ is the energy difference between $\left|e_{1,2}\right>$ and $\left|g\right>$. Under the rotating wave approximation and after an appropriate unitary transformation, the Hamiltonian of the system can be written as \cite{Cohen-Tannoudji1991,Cohen2011,Gaubatz1990} :
\begin{eqnarray}
& & \widehat{H}= \frac{\hbar}{2} \left\{   -  2\delta_1\left|e_1\right>\left<e_1\right| - 2\delta_2\left|e_2\right>\left<e_2\right| \right.  \nonumber\\
& & + \left. \Omega_1\left[ \left|e_1\right>\left<g\right| + \text{h.c.} \right]
 +  \Omega_2\left[ \left|e_2\right>\left<g\right| + \text{h.c.} \right] \right\} \;.
\label{eq_STIRAP_H}
\end{eqnarray}
In the case where the two-photon detuning $\delta\equiv\delta_1-\delta_2$ is zero, one of the eigenstates of $\widehat{H}$ takes the form \cite{Cohen-Tannoudji1991,Cohen2011,Gaubatz1990}~:
\begin{equation}
\left|\Phi_0\right> = \frac{\Omega_2}{\sqrt{\Omega_1^2+\Omega_2^2}} \left|e_1\right> - \frac{\Omega_1}{\sqrt{\Omega_1^2+\Omega_2^2}} \left|e_2\right>  \;,
\label{eq_Dark_State}
\end{equation}
which, remarkably, does not involve state $\left|g\right>$. In order to transfer the population from $\left|e_{1}\right>$ to $\left|e_{2}\right>$ following the usual STIRAP protocol, the microwave field $\Omega_2$ is turned on first (with $\Omega_1=0$), such that $\left|\Phi_0\right> = \left|e_1\right>$ at the beginning of the sequence. Then, $\Omega_2$ is gradually ramped down to zero while $\Omega_1$ is ramped up to its maximum value, such that $\left|\Phi_0\right>$ becomes proportional to $\left|e_2\right>$ at the end of the sequence. If the variations in the Rabi frequencies $\Omega_1$ and $\Omega_2$ are sufficiently slow, the atoms will adiabatically follow the change in $\left|\Phi_0\right>$, and thus be transferred from $\left|e_1\right>$ to $\left|e_2\right>$ without populating the intermediate state $\left|g\right>$. More precisely, the adiabaticity condition reads \cite{Bergmann1998}~: $\dot{\theta}^2 \ll |E_0-E_\pm|^2/\hbar^2$, where $\dot{\theta}=(\dot{\Omega}_1 \Omega_2 - \Omega_1 \dot{\Omega}_2)/(\Omega_1^2+\Omega_2^2)$, $E_0$ is the energy of state $\left|\Phi_0\right>$ and $E_\pm$ are the two other eigenvalues of $\widehat{H}$. For example, in the particular case where $\Omega_1=\Omega_0 \cos[\pi t / (2\tau)]$ and $\Omega_2=\Omega_0 \sin[\pi t / (2\tau)]$ with $0\leq t \leq \tau$ (see the following sections for a possible practical implementation), the adiabaticity condition takes the following simple form~:
\begin{equation} 
\tau \gg 1/(\sqrt{\Omega_0^2+\Delta^2}-|\Delta|)\;,
\label{eq_adiab}
\end{equation}
where $\tau$ is the duration of the STIRAP process. In this simple model, the maximum efficiency is obtained for zero one- and two-photon detunings ($\Delta=\delta=0$), a condition that will change in the following sections when considering a more realistic experimental situation.

\subsection{Modelling losses and collisions} 
\label{section_coll}

Inter-particle interactions play a significant role in our experiment, all the more that we use BECs with relatively high density. These effects include both energy shifts and collisional losses \cite{Dalfovo1999}. In the following, we will take these effects (and other loss mechanisms) into account using the formalism of non-linear Bloch equations.

Following \cite{Shore1990}, we model the effects of surroundings upon our three-level system by introducing an ensemble of additional states called a bath, and by considering the reduced density matrix $\widehat{\rho}$ obtained by tracing over bath variables. Throughout this paper, we will use the standard notation $\rho_{lj}=\left<l\right|\widehat{\rho} \left|j\right>$, with $l,j\in\{e_1,g,e_2\}$. In the absence of loss and collisions, the evolution of $\widehat\rho$ is described by $i\hbar \partial \widehat{\rho} / \partial t = [\widehat{H},\widehat{\rho}]$.

The effects of atomic losses due to the anti-trapping behaviour of state $\left|g\right>$ is modelled by a constant loss rate $\Gamma$. We neglected collisional effects for atoms in state $\left|g\right>$ because the population of the latter remains very low at all times. Atoms confined in the two other states $\left|e_1\right>$ and $\left|e_2\right>$ are subject to collisional effects. In the mean field theory, these can be modeled \cite{Dalfovo1999} by adding to the Hamiltonian (\ref{eq_STIRAP_H}) a term of the form~: $n_0 g (\rho_{e_1e_1}\left|e_1\right>\left<e_1\right|+\rho_{e_2e_2}\left|e_2\right>\left<e_2\right|)$, where $n_0\sim$~5$\cdot$10$^{14}$~cm$^{-3}$ is the typical density of our BEC, and $g$ is the coupling constant related to the scattering length $a$ through $g=4\pi \hbar^2 a/m$, with $m$ the atomic mass and $a\simeq 5.77$~nm for $^{87}$Rb \cite{Dalfovo1999}. Although the typical value of $n_0 g /h$ in our experiment (on the order of a few kHz) is not negligible with respect to other parameters, mean field collisions shift in our model only lead to very small visible effect on the numerical simulations of the STIRAP process that will be described in the following sections.

On the other hand, inelastic collisions of the form $\left|2,1\right>+\left|2,1\right>\rightarrow\left|2,0\right>+\left|2,2\right>$ play a significant role in the dynamics of our system. Note that atoms falling into $\left|2,0\right>$ are untrapped and therefore lost. For this reason, we shall neglect the reverse collision process. Other processes with different output and input states are also neglected, because of conservation rules. The rate of the relevant collision process, which can be rewritten as $\left|e_2\right>+\left|e_2\right>\rightarrow\left|2,0\right>+\left|e_1\right>$, has the form $\Gamma_{col} \rho_{e_2e_2}$, where $\Gamma_{col}=\gamma_{22} n_0$ can be obtained from the value of $\gamma_{22}$ published in the literature (typically, $\gamma_{22} \simeq$~10$^{-19}$~$\text{m}^3.\text{s}^{-1}$ \cite{Egorov2013}, which results in $\Gamma_{col} \simeq $~50~s$^{-1}$ in our case). Following \cite{Shore1990}, we model this collision process by the combination of a loss process on level $\left|e_2\right>$ and a population flow from $\left|e_2\right>$ to $\left|e_1\right>$, both occurring at the rate $\Gamma_{col} \rho_{e_2e_2}$.

This procedure leads \cite{Shore1990} to the following equations for the population terms of the reduced density operator~:
\begin{eqnarray}
\dot{\rho}_{e_1e_1} & = & i\frac{\Omega_1}{2}\left(\rho_{e_1g}-\rho_{ge_1} \right) + \Gamma_{col}\rho^2_{e_2e_2}  \;, \nonumber \\
\dot{\rho}_{gg} & = & i\frac{\Omega_1}{2}\left(\rho_{ge_1}-\rho_{e_1g}\right) + i\frac{\Omega_2}{2}\left(\rho_{ge_2}-\rho_{e_2g}\right) \nonumber \\
& & \qquad -\Gamma\rho_{gg}  \;, \nonumber \\
\dot{\rho}_{e_2e_2} & = & i\frac{\Omega_2}{2}\left(\rho_{e_2g}-\rho_{ge_2} \right) - 2\Gamma_{col}\rho^2_{e_2e_2}  \;,
\label{eq_STIRAP_pop}
\end{eqnarray}
and for the coherences~:
\begin{eqnarray}
\dot{\rho}_{e_1g} & = & i\tilde\delta_1\rho_{e_1g} + i\frac{\Omega_2}{2}\rho_{e_1e_2} + i\frac{\Omega_1}{2}\left(\rho_{e_1e_1}-\rho_{gg}\right)  \nonumber\\
& & \qquad - \frac{\Gamma}{2}\rho_{e_1g}   \;, \nonumber\\
\dot{\rho}_{e_1e_2} & = & i\left(\tilde\delta_1-\tilde\delta_2\right)\rho_{e_1e_2} + i\frac{\Omega_2}{2}\rho_{e_1g} - i\frac{\Omega_1}{2}\rho_{ge_2}    \nonumber\\
& & \qquad - \Gamma_{col}\rho_{e_2e_2}\rho_{e_1e_2}    \;, \nonumber\\
\dot{\rho}_{ge_2} & = & i\frac{\Omega_2}{2}\left(\rho_{gg}-\rho_{e_2e_2}\right) - i\frac{\Omega_1}{2}\rho_{e_1e_2} - i\tilde \delta_2\rho_{ge_2}  \nonumber \\
& & \qquad - \frac{\Gamma}{2}\rho_{ge_2} - \Gamma_{col}\rho_{e_2e_2}\rho_{ge_2} \;,
\label{eq_STIRAP_coh}
\end{eqnarray}
where $\tilde \delta_1=\delta_1-n_0 g \rho_{e_1e_1}$ and $\tilde \delta_2=\delta_2-n_0 g \rho_{e_2e_2}$ include mean field effects. This set of equations will be the starting point for the numerical simulation of our experimental sequence, which will be described in the next sections.

%%%%%%%%%%%%%%%%%%%%%%%%%%%%%%%%%%%%%%%%%%%%%%%%%%%%%%%%%%%%%
%                                                           %
%%%%%%%%%%%%%%%%%%%%%%%%%%%%%%%%%%%%%%%%%%%%%%%%%%%%%%%%%%%%%
\section{Experiment}
\label{sec_experiment}

\subsection{Protocol}
\label{subsec_protocol}

We start using a setup similar to \cite{Farkas2010}, with a $^{87}$Rb BEC of $\left(6,8\pm0,6\right)\cdot10^{3}$ atoms, which is transferred after the evaporative cooling stage into a harmonic trap with trapping frequencies 120~Hz and 210~Hz in the horizontal plane, and 228~Hz in the vertical plane. The STIRAP is performed in the latter trap.

\begin{figure}
\centering \def\svgwidth{0.48\textwidth}
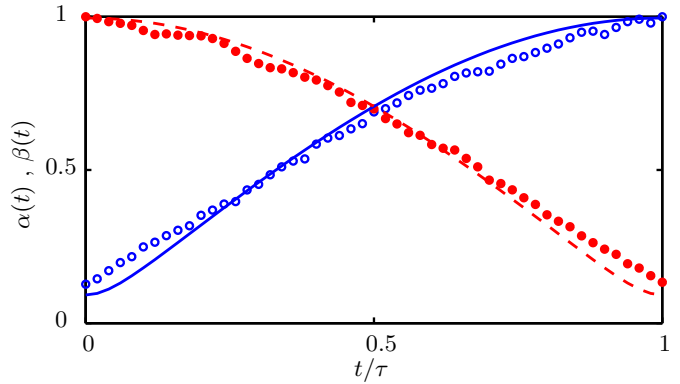
\caption{\label{fig_Stirap_ramp} Relative amplitude of the microwave fields $\alpha$ and $\beta$ used for the STIRAP protocol, as a function of time; solid blue line and open blue circles~: $\alpha \propto \Omega_1(t)$ tuned to the transition between $\left|e_1\right>$ and $\left|g\right>$; red dashed line and filled red circles~: $\beta \propto \Omega_2(t)$ tuned to the transition between $\left|e_2\right>$ and $\left|g\right>$. Circles are the measured values at the output of the amplifier and solid and dashed lines are plotted after the model of equations (\ref{eq_RabiRamps}).}
\end{figure}

The two microwave fields needed for the experimental protocol are generated by I/Q modulation. For this purpose, we use a microwave carrier wave of the form $C \propto \cos(2\pi f_{mw}t)$, and two radio-frequency ``in-phase" and ``quadrature" signal waves of the form $S_{\pm} \propto \cos\left[2\pi f_{rf}t+\varphi_{\pm}(t)\right]$, with
\begin{equation}
\varphi_{\pm}(t) = \pm \left( \frac{\pi t}{2 \tau} - \frac{\pi}{4} \right) \;,
\label{eq_phase_mod}
\end{equation}
where $t \in \left[0,\tau\right]$. This results in a microwave with magnetic field of the form $\mathbf{B_{mw}} \propto (B_{mw1}+B_{mw2})\mathbf{u}$, where $\mathbf{u}$ is a unit (linear) polarization vector and~:
\begin{eqnarray}
B_{mw1}(t) & \propto & B_{mw} \alpha(t) \cos[2\pi(f_{mw}+f_{rf})t]  \;, \nonumber\\
B_{mw2}(t) & \propto & B_{mw} \beta(t) \cos[2\pi(f_{mw}-f_{rf})t]  \;,
\label{eq_RabiRamps}
\end{eqnarray}
corresponding to two side bands $f_{mw}+f_{rf}$ and $f_{mw}-f_{rf}$ with their relative amplitudes $\alpha(t)=\sin[\pi t/(2\tau)]$ and $\beta(t)=\cos[\pi t/(2\tau)]$ changing over time. In order to take into account the non-zero extinction ratio $\epsilon$ of our modulator (typically $\epsilon = 2\%$ in our case), we introduce the following phenomenological expressions for $\alpha$ and $\beta$~:
\begin{eqnarray}
{\alpha(t)} & \propto & \sqrt{(1-\epsilon)\sin^2\left( \frac{\pi t}{2\tau} \right) + \epsilon}  \;,  \nonumber\\
{\beta(t)} & \propto & \sqrt{(1-\epsilon)\cos^2\left( \frac{\pi t}{2\tau} \right) + \epsilon}  \;.
\label{eq_RabiRamps}
\end{eqnarray}
In practice, $\tau$ is the duration of the STIRAP sequence, $\omega_1$ and $\omega_2$ are respectively equal to $2\pi(f_{mw}+f_{rf})$ and $2\pi(f_{mw}-f_{rf})$, and $\Omega_1$ and $\Omega_2$ are proportional to $\alpha$ and $\beta$. The overall microwave signal is amplified to 40~dBm before being radiated by a microwave horn in the direction of the atomic cloud. The modelled variations of $\alpha$ and $\beta$ are shown by solid and dashed lines in figure \ref{fig_Stirap_ramp}. The corresponding experimental power levels, measured at the output of the microwave amplifier, are plotted as circles in the same figure, showing a relatively good agreement.

Since $\omega_1$ is mostly resonant with the transition $\left|e_1\right>\leftrightarrow\left|g\right>$ and $\omega_2$ is mostly resonant with the transition $\left|e_2\right>\leftrightarrow\left|g\right>$, the Rabi frequencies $\Omega_{1,2}$ can be written as $\hbar\Omega_1=\sqrt{3/4} \mu_B |B_\bot|$ and $\hbar\Omega_2=\sqrt{3/4} \mu_B |B_\||$, where $\mu_B$ is the Bohr magneton, $B_\bot = B_{mw1}(\mathbf{u} - \mathbf{u} \cdot \mathbf{B}/|\mathbf{B}|)$ is the component of $B_{mw1}\mathbf{u}$ in the plane perpendicular to the local direction of the DC magnetic field $\mathbf{B}$, and $B_\| = B_{mw2}\mathbf{u} \cdot \mathbf{B}/|\mathbf{B}|$ is the component of $B_{mw2}\mathbf{u}$ along to the local direction of $\mathbf{B}$. For a given amount of microwave power, the STIRAP duration $\tau$ (which must be longer than $1/\Omega_0$ to satisfy the adiabaticity condition) is minimized for $\max\left[ \Omega_1(t) \right] = \max\left[ \Omega_2(t) \right] $, which is expected to occur in our experimental setup for $\max|B_\bot|= \max|B_\||$, corresponding to an angle of 45$^\circ$ between $\mathbf{u}$ and $\mathbf{B}$. In practice we also aim for $\mathbf{u}$ to be almost parallel to the plane of the atom chip (which is compatible with the latter condition), as it contains several conductive wires (and considering the fact that in the case of a perfectly conducting plane the magnetic field would have to be parallel to the latter). Experimentally, we obtained $\Omega_1(t=\tau)\simeq43$~kHz and $\Omega_2(t=0)\simeq14$~kHz (see section \ref{sec_Simulation} for the estimation procedure). We attribute the difference between these two values to residual uncertainties on the orientation of the horn, the magnetic field at the bottom of the trap and the propagation of the microwave field from the horn to the atoms. Still, the two values are sufficiently close to perform an efficient STIRAP protocol, as will be shown in the next section.

To distinguish between atoms in states $\left|e_1\right>$ and $\left|e_2\right>$ and measure the transfer efficiency, we let the atoms fall under gravity in a magnetic field gradient after the STIRAP sequence, which results in their spatial separation as shown in figure \ref{fig_atomic_clouds}.

\subsection{Results}

\begin{figure}
\centering \def\svgwidth{0.48\textwidth}
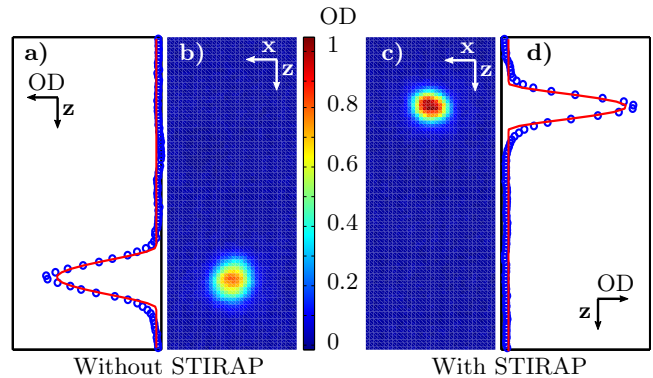
\caption{\label{fig_atomic_clouds} a) vertical optical density (OD) profile of the BEC after time-of-flight in $\left|F=2,m_F=2\right>$ without the STIRAP and b) optical density of the same BEC. c) optical density profile of the BEC after time-of-flight in $\left|F=2,m_F=1\right>$ with a 900~$\mu$s STIRAP sequence and d) vertical optical density profile of the same BEC in $\left|F=2,m_F=1\right>$. Blue open circle stand for experimental data, and the solid red line is a parabolic fit. In order to discriminate between the Zeeman sublevels of the $F=2$ manifold, during the time-of-flight we apply a magnetic field gradient in the vertical direction of the plot. This gradient causes different accelerations for atoms with different $m_F$ numbers resulting in different spatial positions. Without the STIRAP we produce a BEC of $\left(6,8\pm0,6\right)\cdot10^{3}$ atoms and with the STIRAP sequence we produces a BEC of $\left(6,0\pm0,5\right)\cdot10^{3}$ atoms. The transfer efficiency is around $87\%$. }
\end{figure}

In order to determine the optimal values for the microwave frequencies, we first measured the modulus of the magnetic field at the bottom of the trap with radio-frequency spectroscopy. We found 2,45~G~$\pm$~7~mG, which led, based on the Breit-Rabi formula \cite{Steck2003}, to the following target values~: $f_{rf}~=$~856,3~$\pm$~2,4~kHz and $f_{mw}~=$~6,838~976~GHz~$\pm$~12,3~kHz to match the one- and two-photon resonance conditions. We then varied the three experimental parameters $f_{rf}$, $f_{mw}$ and $\tau$ to maximize the transfer efficiency. The optimal parameters were found to be~: $\tau=$~900~$\mu$s, $f_{rf}=$~860,5~kHz and $f_{mw}=$~6,838~945~GHz. This corresponds to a transfer efficiency around $87\%$, as illustrated in figure \ref{fig_atomic_clouds} which shows the optical density of a BEC without and with the STIRAP sequence. A summary of the atomic and field frequencies used in the optimal case is provided in table \ref{tab_freq_stirap}. Interestingly, they correspond to non-zero values of the one- and two-photon detunings, a point which will be discussed in detail in the following sections.

\begin{table}%add [H] placement to break table across pages
\caption{\label{tab_freq_stirap} Atomic transition frequencies $\omega_{01,2}$ computed using the Breit Rabi formula \cite{Steck2003} and the measured magnetic field value of 2,446~G (obtained from the numerical fit, see section \ref{sec_Simulation}), and microwave frequencies $\omega_{1,2}$ experimentally obtained by optimizing the STIRAP efficiency. We deduce the values of the two-photon detuning $\delta /(2\pi)=$~11~kHz and the one-photon detuning $\Delta /(2\pi) =$~-24,5~kHz.}
\begin{ruledtabular}
\begin{tabular}{ccc}
 transition~: & $\left|e_1\right>\leftrightarrow\left|g\right>$ & $\left|e_2\right>\leftrightarrow\left|g\right>$ \\
 $\omega_{0i}/(2\pi)$ $[\text{GHz}]$~: & 6,839~824~5 & 6,838~114~5 \\
 $\omega_i/(2\pi)$ $[\text{GHz}]$~: & 6,839~805~5 & 6,838~084~5 \\
 $\delta_i/(2\pi)$ $[\text{kHz}]$~: & -19  & -30 \\
\end{tabular}
\end{ruledtabular}
\end{table}
From the data in figure \ref{fig_atomic_clouds}, we observe that the maximum optical density after the population transfer is higher. We attribute this to an oscillation in the size of the BEC, as a consequence of the rapid change (i.e. $\tau$ lower than the inverse of the trap frequencies) in the magnetic potential during the STIRAP process.

\begin{figure}
\centering \def\svgwidth{0.48\textwidth}
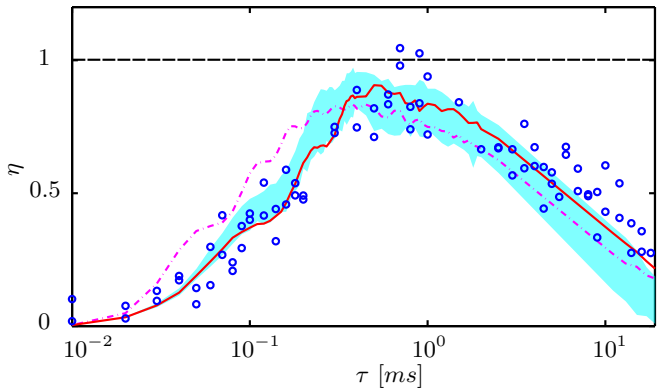
\caption{\label{fig_EtaTau} Transfer efficiency $\eta$ as a function of the duration of the STIRAP sequence for $f_{mw}=$~6,838~945~GHz and $f_{rf}=$~860,5~kHz (optimal parameters). Open blue circles~: experimental data. Solid red line~: model of equations (\ref{eq_STIRAP_pop}) and (\ref{eq_STIRAP_coh}) with the fitted parameters given in section \ref{sec_Simulation}. Cyan surface~: simulation with the fitted parameters and with noise as discussed in section \ref{sec_Simulation}. Dash-dot magenta line~: model of equations (\ref{eq_STIRAP_pop}) and (\ref{eq_STIRAP_coh}) with the fitted parameters and setting the one- and two-photon detunings to zero.}
\end{figure}

\begin{figure*}
\centering \def\svgwidth{1\textwidth}
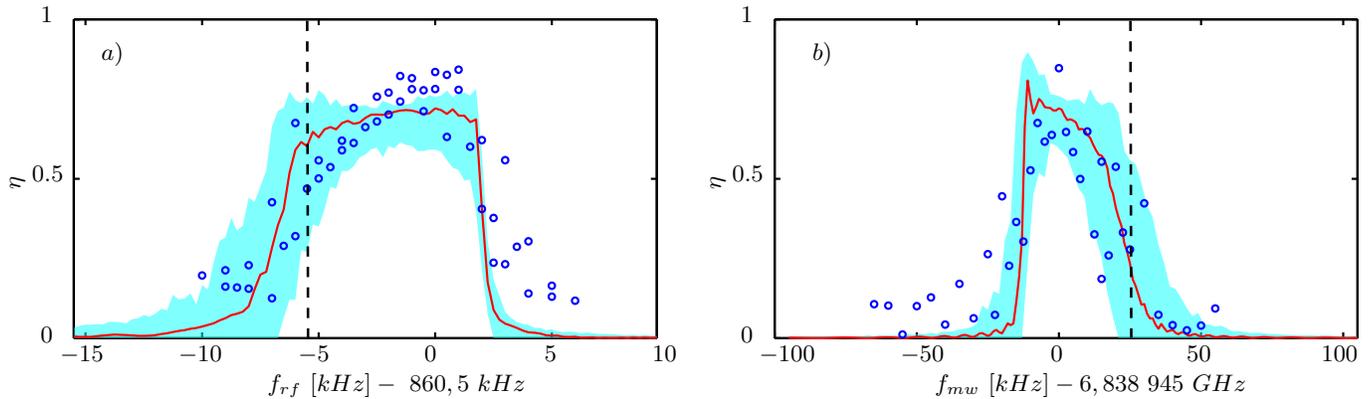
\caption{\label{fig_EtaF}  a)~: variation of the transfer efficiency $\eta$ as a function of $f_{rf}$ (which is related to the two-photon detuning $\delta$ by $\delta=4\pi f_{rf} + \omega_{02}-\omega_{01}$), for $\tau=$~2,5~ms and $f_{mw}=$~6,838~945~GHz. b)~: variation of the transfer efficiency $\eta$ as a function of $f_{mw}$ (which is related to the one-photon detuning $\Delta$ by $\Delta=2\pi f_{mw} - (\omega_{01}+\omega_{02})/2$), for $\tau=$~2,5~ms and $f_{rf}=$~860,5~kHz. Open blue circles~: experimental data, solid red line~: model of equations (\ref{eq_STIRAP_pop}) and (\ref{eq_STIRAP_coh}). Cyan surface~: simulation with the fitted parameters and with noise as discussed in section \ref{sec_Simulation}. In both sets of curve the frequency offset corresponds to the experimental parameters of figure \ref{fig_atomic_clouds} (optimal parameters). The vertical black dashed lines are the positions of the one- and two-photon resonances theoretically estimated from the fitted value of the magnetic field $B_0=$~2,446~G. }
\end{figure*}

For the optimal frequencies, we plot in figure \ref{fig_EtaTau} (open circles) the experimental transfer efficiency against the duration of the frequency ramp. First the efficiency increases to near 100~\% in 900~$\mu$s, and then it exponentially decreases to near 0~\% at a rate of 60~s${}^{-1}$. For $\tau=$~2,5~ms \footnote{The data in figure \ref{fig_EtaF} was taken before we were able to create short STIRAP pulses, thus in figure \ref{fig_EtaF} $\tau$ is longer than the optimal value in figure \ref{fig_EtaTau}.} we plot in figure \ref{fig_EtaF} (open circles) the one- and two-photon resonance curves, obtained by varying respectively $f_{mw}$ (figure \ref{fig_EtaF}.b) and $f_{rf}$ (figure \ref{fig_EtaF}.a). Experimentally we found a 45~kHz linewidth for the one-photon resonance curve and a 18~kHz linewidth for the two-photon resonance curve.

The dispersion in the experimental points mostly corresponds to the shot to shot fluctuations in the atom number before the STIRAP sequence.

%%%%%%%%%%%%%%%%%%%%%%%%%%%%%%%%%%%%%%%%%%%%%%%%%%%%%%%%%%%%%
%                                                           %
%%%%%%%%%%%%%%%%%%%%%%%%%%%%%%%%%%%%%%%%%%%%%%%%%%%%%%%%%%%%%
\section{Simulation}
\label{sec_Simulation}

\subsection{Simulation protocol}

To simulate the behaviour of our STIRAP protocol, we numerically integrate equations (\ref{eq_STIRAP_pop}) and (\ref{eq_STIRAP_coh}), with the Rabi frequency ramps of equation (\ref{eq_RabiRamps}). This set of equations contains five independent parameters~: the magnetic field modulus at the position of the atoms $B_0=|\mathbf{B}|$ which determines the two detunings $\delta_1$ and $\delta_2$, the maximum Rabi frequencies for the two microwave ramps $\Omega_1(t=\tau)$ and $\Omega_2(t=0)$, and the loss rates $\Gamma_{col}$ and $\Gamma$. A fit to the three experimental data sets shown in figures \ref{fig_EtaTau}, \ref{fig_EtaF}.a and \ref{fig_EtaF}.b was performed to estimate these five parameters. It led to~: $\Omega_1(t=\tau)=2\pi\times$43,4~kHz, $\Omega_2(t=0)=2\pi\times$14,4~kHz, $B_0=$~2,446~G, $\Gamma=$~635~s$^{-1}$ and $\Gamma_{col}=$~160~s$^{-1}$. The corresponding fitting curves are plotted as red lines in figures \ref{fig_EtaTau}, \ref{fig_EtaF}.a and \ref{fig_EtaF}.b.

The oscillations of the simulated transfer efficiency with $\tau$ (see red solid line in figure \ref{fig_EtaTau}) are attributed to the fact that $\Omega_2(\tau)$ and $\Omega_1(0)$ are not exactly zero as they ideally should be (see equations (\ref{eq_RabiRamps})). This induces a non-zero population of the three eigenstates of $\widehat{H}$, which interfere and give rise to these oscillations.

In order to account for the noise in the current sources and ambient magnetic field (our experiment was performed without any magnetic shielding), we added in the simulation a magnetic Gaussian white noise, with a standard deviation of 2,5~mG in the time domain and an effective frequency range between 4,5~Hz and 450~kHz. About one hundred independent runs of the simulation were performed, with random realisations of the magnetic field noise. In figures \ref{fig_EtaTau} and \ref{fig_EtaF}, the cyan surface show the two-standard-deviation dispersion of the transfer efficiency around its mean value. This allows us to capture some of the dispersion of our experimental data.

\subsection{Discussion of the fitted parameters}
\label{subsec_STIRAP_line_shape}

The fitted value for $\Omega_1(t=\tau)$ is in good agreement with an independent measurement of 43~kHz~$\pm$~5~kHz obtained by direct Rabi oscillations on the transition $\left|e_1\right>\leftrightarrow\left|g\right>$. The value of $B_0$ is also in good agreement with the magnetic field modulus 2,45~G~$\pm$~7~mG measured by radio-frequency spectroscopy. The values of $\Omega_1(t=\tau)$ and $\Omega_2(t=0)$ are not equal for the reasons already discussed in section \ref{subsec_protocol}.

The fitted value of $\Gamma_{col}$ is of the same order of magnitude as the estimated value of section \ref{section_coll}. For $\Gamma$, it is not straightforward to give a theoretical estimate as it results from multiple and complex loss mechanisms. However, we observe that the order of magnitude of the fitted value (635~s$^{-1}$) is consistent with the inverse of $t_{out}$ ($1.4 \times 10^3$~s$^{-1}$), defined as the time taken for an atom initially at rest to fall under gravity by a distance $\sigma_z=$~2,3~$\mu$m, which is half the vertical dimension of the BEC (obtained using the Thomas-Fermi approximation). Note that the repulsive magnetic force has in our case the same order of magnitude as the gravity force, hence it does not change the order of magnitude for $t_{out}$.

To conclude, our numerical model including losses and collisions shows a very good agreement with the experimental data, with realistic values of the fitted parameters.

\subsection{Numerical simulation at one- and two-photon resonance}

In order to emphasize the importance of having non-zero values of both the one-photon ($\Delta$) and two-photon ($\delta$) detunings, we show in figure \ref{fig_EtaTau} (dash-dot magenta line) a numerical simulation of the transfer efficiency versus $\tau$ using the fitted parameters discussed above, with the additional condition $\delta=\Delta=0$. As expected, the overall efficiency of the STIRAP process is lower in this case. In the next sections, we will give some physical insights on the reason why non-zero detunings improve the situation in this particular case.

%%%%%%%%%%%%%%%%%%%%%%%%%%%%%%%%%%%%%%%%%%%%%%%%%%%%%%%%%%%%%
%                                                           %
%%%%%%%%%%%%%%%%%%%%%%%%%%%%%%%%%%%%%%%%%%%%%%%%%%%%%%%%%%%%%
\section{Physical discussion: why do we need non-zero one- and two-photon detunings?}
\label{sec_discution}

\subsection{Effect of collision losses}

To make the physical effect of collision losses more obvious, we will use in the following a much simpler theoretical model than non-linear Bloch equations. For this purpose, we consider only a pure state $\left|\psi(t) \right>$, and we write the number of atoms in state $\left|e_2\right>$ as $N_2(t)=|\left<e_2|\psi(t)\right>|^2 N(t)$, where $N(t)$ is the overall atom number \cite{Romanenko1997}. In this framework, the instantaneous loss rate for $N_2$ due to the inelastic collisions in $\left|e_2\right>$ is $\Gamma_{col} |\left<e_2|\psi(t)\right>|^2$, leading to the following equation for $N(t)$~:
\begin{equation} \label{eq_de_N}
\dot{N}=-\Gamma_{col} |\left<e_2|\psi(t)\right>|^4 N \;.
\end{equation}
In an ideal STIRAP sequence, $\left|\psi(t)\right>$ is equal at any time to the eigenstate $\left|\Phi_0(t)\right>$ (given by equation (\ref{eq_Dark_State})) of $\widehat{H}_0$, the latter being defined as the Hamiltonian of equation (\ref{eq_STIRAP_H}) with the additional condition $\delta=0$. To go further, we assume that the two-photon detuning $\delta$ is small enough to be treated as a perturbation of $\widehat{H}_0$, which is the case if $\left|\delta\right|$ is much smaller than all the differences between the eigenvalues of $\widehat{H}_0$. This condition can be rewritten as~:
\begin{equation} \label{eq_perturb}
\frac{|\delta|}{\sqrt{\Omega_0^2 + \Delta^2}-|\Delta|} \ll 1 \;,
\end{equation}
where $\Omega_0^2=\Omega_1^2 + \Omega_2^2$. Applying the perturbation theory to $\left|\Phi_0\right>$ up to the first order in the small parameter defined in equation (\ref{eq_perturb}) then leads to a corrected state $|\Phi_0^{(corr)}>$, with the following expression for $|<e_2|\Phi_0^{(corr)}>|^2$~:
\begin{equation}
\left|\left<e_2|\Phi_0^{(corr)}\right> \right|^2=\frac{\Omega_1^2}{\Omega_0^2} \left( 1+\frac{8 \delta \Delta \Omega_2^2}{\Omega_0^4} \right)\;.
\end{equation}
Inserting the latter equation in (\ref{eq_de_N}) and integrating over time by assuming ideal STIRAP pulses of the form $\Omega_1=\Omega_0 \cos(\pi t/(2\tau))$ and $\Omega_2=\Omega_0 \sin(\pi t/(2\tau))$ yields the following expression for $N_2$ at the end of the STIRAP sequence~:
\begin{equation} \label{eq_pertes_col}
N_2(\tau)= \exp \left[-\Gamma_{col} \tau \left(\frac{3}{8}+\frac{\delta \Delta}{ \Omega_0^2} \right) \right] N(0)\;.
\end{equation}
As can be seen in this equation, the effect of collision losses can be mitigated by choosing a non-zero value for both $\delta$ and $\Delta$. For this purpose, the product $\delta \Delta$ has to be negative, which is indeed what we found experimentally when optimizing the parameters for maximum transfer efficiency (see table \ref{tab_freq_stirap} and related caption). Physically, the reduction in losses can be understood as resulting from the slower growth of the population in $\left|e_2\right>$ during the STIRAP sequence, reducing the instantaneous loss rate $\Gamma_{col}|\left<e_2|\psi(t)\right>|^2$ (at the price of a more stringent adiabaticity condition (\ref{eq_adiab})). In practice, $|\delta|$ cannot be made too big because the effect described above is counterbalanced by losses from the anti-trapped state $\left|g\right>$, which can be shown under the same hypotheses to induce additional losses on $N_2$ of the form $\exp[-\delta^2 \Gamma \tau/(2\Omega_0^2)]$ \cite{Romanenko1997}.

In conclusion, we have shown in this section that the need for non-zero one- and two-photon detunings could be understand as a consequence of the collision-induced losses in $\left|e_2\right>$. It should be kept in mind however that equation (\ref{eq_pertes_col}) is only valid for small values of $|\delta|$, and cannot quantitatively predict the optimal values of the detunings in our case (although it correctly predicts the sign of $\delta \Delta$). 

\subsection{Effect of the imbalance of the Rabi frequencies}

\begin{figure*}
\centering \def\svgwidth{1\textwidth}
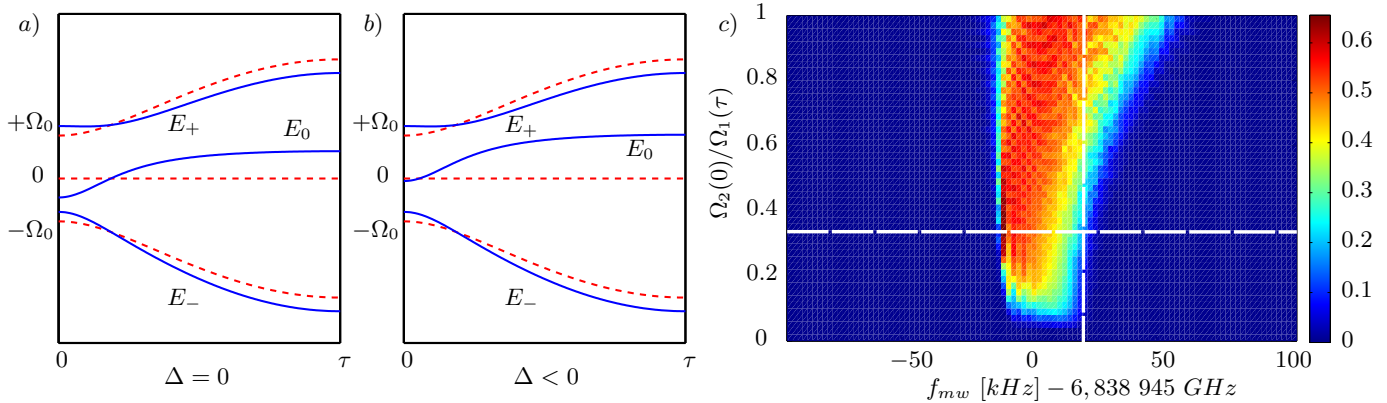
\caption{\label{fig_EtaAlpha2} a)~:  variations of the eigen-energies as a function of time (equations (\ref{eq_AdiabaticEnergies})), in the case $\delta=$~10~kHz, $\Omega_2(0)=$~14,4~kHz and $\Omega_1(\tau)=$~43,4~kHz, for $\Delta=$~0; b)~: same curves with $\Delta$=-6~kHz; c)~: transfer efficiency (full simulation with fitted parameters) as a function of the microwave frequency $f_{mw}$ and the ratio of the maximum Rabi frequencies $\Omega_2(0)/\Omega_1(\tau)$. The vertical white dashed line indicates $\Delta=$~0. The horizontal white dashed line indicates the ratio $\Omega_2(0)/\Omega_1(\tau)$ used in our experiment.}
\end{figure*}

In this section, we will show how the need for a non-zero one-photon detuning $\Delta$ is accentuated by the fact that the maximum values of the two Rabi frequencies $\Omega_1(\tau)$ and $\Omega_2(0)$ are not equal. For this purpose, we proceed in the same way as for the previous section, with the additional hypothesis $|\Delta|\ll \Omega_0$. This leads to the following eigenenergies for $\widehat{H}+\hbar \Delta$ up to the first order in $|\delta|/\Omega_0$ and $|\Delta|/\Omega_0$~:
\begin{eqnarray}
E_0 & = & \frac{\hbar}{2}\left[ - \Delta - \delta\frac{\Omega_2^2-\Omega_1^2}{\Omega_0^2} \right] \;, \nonumber \\
E_+ & = & \frac{\hbar}{2}\left[ + \Omega_0 + \delta\frac{\Omega_2^2-\Omega_1^2}{2\Omega_0^2} \right] \;, \nonumber \\
E_- & = & \frac{\hbar}{2}\left[ - \Omega_0 + \delta\frac{\Omega_2^2-\Omega_1^2}{2\Omega_0^2} \right] \;.
\label{eq_AdiabaticEnergies}
\end{eqnarray}
The shape of these three energy curves versus time when $\Delta=0$ is shown in figure \ref{fig_EtaAlpha2}.a, where they have been plotted using Rabi frequency ramps of the form~$\Omega_1=\Omega_1(\tau) \cos(\pi t/(2\tau))$ and $\Omega_2=\Omega_2(0) \sin(\pi t/(2\tau))$, with the following values~: $\delta=$~10~kHz, $\Omega_1(\tau)=$~43,4~kHz and $\Omega_2(0)=$~14,4~kHz. The parameters used to plot the curves slightly violate the conditions $|\delta|/\Omega_0 \ll 1$ and $|\Delta|/\Omega_0 \ll 1$ to emphasize the effect on the shape of the curves. With $\Omega_2(0)<\Omega_1(\tau)$ as it is the case in our experiment, the adiabaticity condition ($\Omega_1(\tau) \Omega_2(0) / [\tau \Omega_0(t)^2] \ll |E_0-E_{\pm}|/\hbar$) is more stringent around $t=0$. As can be seen in figure \ref{fig_EtaAlpha2}.b, a non-zero and negative value for $\Delta$ relaxes this condition by increasing the minimal distance between the energy curves around $t=0$, making the STIRAP process more efficient. 

A more quantitative study on the interplay between the optimal value of $\Delta$ and the Rabi frequency imbalance, based on the full simulation of our model, is shown in figure \ref{fig_EtaAlpha2}.c, where the transfer efficiency is plotted as a function of $f_{mw}$ and $\Omega_2(0)/\Omega_1(\tau)$, for $\delta=2\pi\times$11~kHz. It can be seen, as expected, that the optimal value for $|\Delta|$ increases as $\Omega_2(0)/\Omega_1(\tau)$ decreases.  

Unlike the shift in the microwave frequencies due to collisional losses, the shift due to imbalance of Rabi frequencies can be cancelled if $\Omega_1(\tau)$ and $\Omega_2(0)$ are made equal. In this case, the full simulation, upper row in figure \ref{fig_EtaAlpha2}.c, shows that the transfer efficiency is still maximum with $\delta\Delta < 0$ as expected. 

%%%%%%%%%%%%%%%%%%%%%%%%%%%%%%%%%%%%%%%%%%%%%%%%%%%%%%%%%%%%%
%                                                           %
%%%%%%%%%%%%%%%%%%%%%%%%%%%%%%%%%%%%%%%%%%%%%%%%%%%%%%%%%%%%%
\section{Conclusion}

We have experimentally demonstrated a microwave STIRAP sequence to transfer a BEC from state $\left|F=2,m_F=2\right>$ to state $\left|F=2,m_F=1\right>$ with a transfer efficiency around $87\%$ in $900\,\mu$s. The STIRAP optimization lead to non-zero values of the one- and two-photon detunings, which can be understood as a result of the mitigation of inelastic collisions which induce losses within $\left|F=2,m_F=1\right>$. The dynamics of our experimental sequence is described to a very good extent by a set of non-linear Bloch equations, with realistic values of the parameters. 

This transfer protocol is a useful tool for atom interferometry with magnetically trapped $^{87}$Rb, as it allows to prepare the atoms in $\left|F=2,m_F=2\right>$, where they have maximal magnetic moment, and coherently transfer them to $\left|F=2,m_F=1\right>$. From this state one can address the $\left|F=2,m_F=1\right>\leftrightarrow\left|F=1,m_F=-1\right>$ two-photon transition, which can be made very robust against magnetically induced decoherence. More generally, this work shows that microwave STIRAP between hyperfine ground states of magnetically trapped BECs is feasible, and quantitatively described by non-linear Bloch equation paving the way for STIRAP-based quantum information or metrology experiments integrated on a chip.

\begin{acknowledgments}
We acknowledge a loan of microwave components and help for the microwave horn design by the Peter Rosenbusch team at SYRTE.
The authors also thank Bruce W. Shore for careful re-reading of their manuscript.
This work has been carried out within the OnACIS project ANR-13-ASTR-0031 funded by the French National Research Agency (ANR) in the frame of its 2013 Astrid program.
\end{acknowledgments}

\bibliography{biblio_these}

\end{document}